\documentstyle[sprocl]{article}

\begin{document}

\title{INTRODUCTION \\
{\bf to the Yuri Golfand Memorial Volume} \\
{\em MANY FACES OF SUPERWORLD} }

\author{M. SHIFMAN}

\address{Theoretical Physics Institute, University of Minnesota,
Minneapolis,
MN 55455}

\maketitle

Supersymmetry is almost thirty years old.
The first supersymmetric model in four dimensions was found by 
Golfand and Likhtman in 1970. Yuri Abramo\-vich Golfand
was a very modest man who, unfortunately, did not gain 
much  recognition  when he was still alive.  
When I saw him
for the last  time  in  Haifa in 1992,
he told me that he had a dream -- to travel to the West,
to the United States or France, to attend a physics conference or just
``to see the world."  This dream never came true.  He died in 1994, but
you will look in vain for the  obituary in {\em Physics Today, CERN 
Courier},
or any Russian physics journals. This Memorial Volume is a belated 
tribute
to the man who was one of the discoverers of supersymmetry,
which today, 30 years later, dominates theoretical
high energy physics. 

\section*{Basic Biographic Data\footnote{Below 
I quote from a note  written by Misha Marinov in 1994
for the {\em  Proceedings of the  Israeli Physical Society},
a source  inaccessible outside Israel. I am grateful to M. Marinov
for providing me with this publication from his archive. Some additional
data were kindly communicated to me by  Mrs. N. Koretz-Golfand.}}

Yuri Abramovich Golfand was born in Kharkov, Ukraine, on January 10, 
1922. Like many  Soviet scholars  of his generation, he started his education at
the Kharkov University, Department of Physics and
Mathematics. This was in 1938. The Second World War (WW II) 
interrupted his academic career; in 1941 Golfand becomes a cadet of  the
Military Airforce Academy. The end of 1944 found him at a front-line
airdrome where he worked as a technician. After the end
of the war, in 1945, Golfand resumed his studies, this time
at the  Department of 
Mathematics of the Leningrad University.
He graduated in 1946 and 
got  his  PhD  in mathematics
 within a year and a half. At the end of the 1940's,
Golfand worked for an electrical engineering research institute. In
1951, he joined the group of I.E. Tamm at the Theory  Department
of the Lebedev Physical Institute (FIAN) in Moscow.
He stayed there for 40 years, with a long break, of which  more will be 
said  later.  For a year or two, Golfand was marginally involved in
 the nuclear bomb project, like many of his colleagues at that time.
Approximately at the same time he got interested in fundamental physics.
 In the 1950's and
60's,  Golfand  carried out several projects in
 quantum field theory, in particular, on 
applications of the functional methods. In 1959, he published a famous
work on the method of renormalization, based on the assumption that
the four-dimensional momentum space has a constant nonzero 
curvature.
That was one of fascinating attempts to introduce elementary length 
to relativistic field theory.

In 1972, the Academy
of Sciences  conducted  a routine campaign of personnel cuts. At the 
FIAN
Theory Department it was decided that Golfand was the least worthy
 member of the group, whose work was unimpactful.
As a result, he
was fired from FIAN in 1973. This unfortunate turn of events
left very little choice to Golfand -- he decides
to  apply for the exit visa to Israel, which only aggravates his situation. 
In due time there comes a refusal.
In those days such an application was considered to be high treason.
Thus, Golfand becomes a {\em refusenik} -- a  nonperson, according to 
the
Orwellian nomenclature --  with all ensuing political
consequences. His struggle  
lasted for  many years. This chapter belongs to a different
book, however, which has yet to be written. We will not touch it.

 Golfand was unemployed for 7 years, until 1980,
when he was accepted  back
to FIAN (but not to the Theoretical  group), 
under  strong pressure from the world physics community, and, in 
particular,
the American Physical Society. 
It was only in June of 1990 -- seventeen years after the
original application -- that the permission was granted
to Golfand's family  to leave
the Soviet Union, which at this time was rapidly approaching its demise.
Within a few months his family moved to Israel. An official
farewell letter from FIAN, signed by Academician L. Keldysh,
the Director General, arrived a few days before the departure.
The concluding paragraph of the
letter reads: ``I would like to express my deep and sincere 
regret for the damage which has been inflicted on you, and henceforth on the 
Institute,
by your dismissal from FIAN." 

It will be fair to add that shortly before this,
the Soviet Academy of Sciences awarded Yuri Golfand with the  
Tamm Prize for Theoretical Physics.  This was the only award Yuri 
Golfand ever received. 

Golfand spent the last years of his life  in Haifa. 
Because of his age, he could not get a regular
professorship, so he settled for a research fellowship at Technion,
under a special program
of the Israeli Government. Yuri Abramovich Golfand died on February 
17, 1994, in Jerusalem, from complications of a brain stroke.

\section*{Work on Supersymmetry; Chronology}

It is known that Golfand discussed the Bose--Fermi  symmetry with his
colleagues in the late 1960's, trying to solve the puzzle of weak
interactions, before the advent of the Glashow-Weinberg-Salam  theory. 
That is  why
he was so much preoccuppied
with the problem of parity violation which is clearly visible 
in the first published work on four-dimensional 
supersymmetry.\cite{one}
Evgeny Likhtman recollects that when he appeared in FIAN as Golfand's 
PhD student,
in the spring 
of 1968,   Golfand had already found an extension of the
Poincar\'e algebra by  bispinor generators. (Today the extension found
by Golfand is referred to as the super-Poincar\'e algebra, while the 
bispinor
generators are called the supercharges.) 
In the review article\cite{2} it is mentioned
that the searches of the extensions
of the Poincar\'e algebra conducted by Golfand in the late 1960's
were originally also motivated
by the desire to bypass the well-known no-go theorems
due to Coleman and Mandula, and Weinberg (or to establish new no-go 
theorems).

Golfand and Likhtman
worked on various aspects of supersymmetry for several years.
Their first published paper\cite{one} entitled ``Extension of the Algebra 
of
Poincar\'e Group  Generators and Violation of $P$
Invariance'' contained
a field-theoretic model, which in modern terms can be described
as supersymmetric quantum electrodynamics (QED) with the mass term 
of the 
photon/photino fields,  plus two chiral matter superfields.
(I suggest we call it  the Golfand--Likhtman model.)
Adding the photon mass term in the Abelian
gauge theory  does not spoil
renormalizability. Alternatively, one can get this
model from massless super-QED by adding a Higgs sector, and 
breaking U(1) spontaneously. The masses of the physical Higgs fields are
then sent to infinity while the photon/photino mass is kept
finite. The requirement of renormalizability was very important
to Golfand who tried to follow as close as
possible the pattern of the only respectable field theory of the time,
quantum electrodynamics. On the other hand, the absence of massless
particles was also a precondition  -- otherwise Golfand and Likhtman
would have had settled for massless super-QED, which is significantly 
simpler
than the model they found.
This  shows that Golfand kept in mind
phenomenological applications in weak interactions.

 The paper\cite{one}  was received by the Editorial office 
of JETP Letters on March 10,
1971. To set the time scale, I should mention that the famous paper
of Gervais and Sakita,\cite{two} known to everybody,  was received
by the Editorial Office of Nuclear Physics on August 13, 1971. Golfand 
and Likhtman
also prepared a detailed publication, which
 appeared in the I.E. Tamm
Memorial Volume.\cite{three}  The only date one can infer now 
with certainty
in connection with this publication
is that this Volume
was sent to print on March 20, 1972. In fact, according to Likhtman's 
memoirs,
 both
papers were prepared practically simultaneously in the end of
1970. For  Western readers I should explain some essential details
regarding the publication process in the Soviet Union. To publish a
scientific paper was much more than just typing the manuscript and
mailing it to the publisher.  There was a long latent period, associated
with getting all sorts of clearances. First, the so-called Expert
Commission (a group of authorized fellow physicists in the given
institution) was supposed to study the paper and recommend its
publication. According to the official rules they had to certify that
 no new discoveries  were reported, because
if they were, the Expert Commission had to recommend to classify the
paper right away. Of course, people tended to stretch the official rules,
otherwise not a single breakthrough  paper would have ever appeared 
in the Soviet Union.

At the next stage the paper would go to the so called Regime
Department whose task was to check that no references to classified
work or undesirable persons were made, no subversive ideas put 
forward,
and so on. With all this paperwork done, the decision to allow
(disallow) publication was to be made by the Director
of the Institute. This is not the
end of the story, however. All materials  intended for publication
had to be cleared through the
so-called GLAVLIT, the almighty agency whose sole obligation was to
ensure  total Censorship in the country. 
If, at the previous stages the author would
have at least some minimal control over what was going on with his
(her) paper, GLAVLIT was a total black box. 

The process of getting all
clearances could extend anywhere from weeks to 
many months, and the paper
was officially nonexistent until the very end. The author could not
even refer to it in his/her further work. Thus, the Likhtman's 
recollections that
the paper\cite{one} was completed in 1970, and the
official submission date of March 10, 1971, are not inconsistent. 

In their second paper on supersymmetry,\cite{three}
Golfand and Likhtman described in detail a recursive procedure
of building supersymmetric models. By this time Likhtman, following 
Golfand's instructions,  worked out 
the free field representations of the super-Poicar\'e algebra 
in several practically important cases (today we would say,
 the chiral and vector supermultiplets were constructed). So, they knew 
how
the numbers of the boson and fermion degrees of freedom
in the supersymmetric Lagrangians should be balanced.
This determined a  starting point -- the particle content of the
models to be built.
Then they suggested cataloguing
all possible interaction terms in the Lagrangian, compatible with 
renormalizability,
order by order in the coupling constant, and the corresponding terms
in the supercharges, with  unknown coefficients.
The coefficients were to be fixed by imposing the anti-commutation 
relation
$\{Q_\alpha \bar Q_{\dot\alpha} \}= 2 P_{ \alpha \dot\alpha }$,
order by order in the coupling
constant. (I use here the modern notation, $Q_\alpha$
denotes the supercharge and $ P_{ \alpha \dot\alpha }$ the energy-
momentum
operator.)
 Needless to say that this was much more 
time-
and labor consuming procedure than the superfield formalism of the 
present
day. Note, however, that the 
 work I describe now took place eight years before the invention
of this formalism. 

In the very same paper,  in addition to the
already established super-Poincar\'e algebra, Golfand and Likhtman
presented
(a limiting case of) the super-deSitter algebra.

Golfand continued to work on this range of ideas even after his
forced retirement, through the years of unemployment.
Misha Marinov recollects:
``Soon after the Wess--Zumino preprint appeared, in January or 
February of 
1974,  I was invited  to
give a talk on supersymmetry  at the Institute of Physical Problems.
Golfand was in the first row and
listened to my explanations very attentively. Then we talked about all
details. Yuri Abramovich was greatly impressed by the Wess--Zumino
work, though he said  it was too technically complicated and that
his approach was more elegant. It is curious to note that
Abrikosov who attended this seminar too, strongly
objected against the exploitation of the prefix ``super'' since it was
already in  use in another context in superconductors.''

Later Golfand, together with Likhtman, wrote an extended 
review\cite{2}
for the collection {\em Supersymmetry: A Decade of Development}, 
edited by Peter West, 
where they summarized their own results and tried to indicate
where they stood in relation to other numerous results
on supersymmetry which were obtained by that time.
This was the last paper written as a team by Golfand and  Likhtman.

\section*{Likhtman}

Under the spell of Golfand's ill fate, the academic path of Evgeny
Pinkhasovich Likhtman went astray. You will read this story
in his memoirs published in this Volume. It should only be added
that in the beginning of the 1970's, before Wess and Zumino,  
Likhtman published several
papers of his own\cite{five,six}, devoted to various aspects of 
supersymmetry. In particular, on page 8 of Ref. \cite{five}   
 one  reads:
``As is known, in relativistic
quantum field theory, in transforming the free energy operator to the
normal-ordered form there emerges an infinite term which is
interpreted as the vacuum energy. It is also known that the sign of
this term is different for  particles subject to the Bose and Fermi
statistics.  The number of  boson states is always equal to the
number of  fermion states. From this it follows that the infinite
positive energy of the boson states in any of the representations of
the [super-Poincar\'e] algebra is annihilated by the infinite 
negative energy of the
fermion states." In one of his JETP Letters papers,\cite{six} Likhtman 
mentions 
in passing
 that in supersymmetric theories the one-loop
boson mass diverges not quadratically, but, like the fermion mass, only
logarithmically. Thus, he apparently
was the first to establish two fundamental
properties of the supersymmetric theories,
distinguishing them from all others
 -- the vanishing of the vacuum energy and the absence
of the quadratic divergences. It was Likhtman who 
 gave a talk on
supersymmetry at ITEP in the 1970's. This was my only personal 
encounter with him.

E. Likhtman remains the employee
of the Institute of  Scientific and Technical Information
in Moscow  till the present.
The only change is that back in the 1970's
the institute was referred to as ``All-Union'',
while now, with the fall of the Soviet Union, this part of the title is gone.

\section*{Missed Crossroads}

It is natural to ask why the ideas of supersymmetry did
not take root in the Moscow particle physics community right away, 
immediately 
after the discovery of Golfand and Likhtman.
The community was strong, vibrant and versatile, and yet it missed
a key turn on the pathway of theoretical physics.

Of course, it is hardly possible now to give a certain answer.
I will still  suggest a few conjectures.

One of the reasons might be a  negative attitude to field theory in 
general
which was prevalent in the community after Landau's discovery of the
``zero charge."
Even after the renormalizability
of the Glashow-Weinberg-Salam model was proved by 't Hooft in 1972,
some of the elders of the community, whose opinions were highly
respected, continued to openly express their
animosity towards field theory. A radical turn occurred only
in November of 1974, after the discovery of $J/\psi$,
with  the advent of quantum chromodynamics (QCD).

Perhaps, more importantly, Golfand was
not taken seriously by many of  his former colleagues. 
To this day some of them
insist that   
``he himself did not understand  
what he did because he was not  really a good physicist.''
This is a quotation from a letter which I got about half a year ago,
when the work on this Volume 
began. The author of the letter then continued:
``I cannot remember a single interesting statement on physics  
which he ever made. Usually he was quite ironic about doing physics.
He would occasionally come to a  
seminar, sit there and then disappear, without saying much or producing  
anything.''  I hasten to add that this  opinion
 is by no means shared by all of Golfand's colleagues. Human
memory is rather selective, and in many cases we see what we want to 
see...
Still, it gives an idea of the general attitude.
  
The point of view 
that Golfand  did not understand  
what he was doing 
is absolutely unsubstantiated either by the analysis
of three papers on supersymmetry produced by Golfand and Likhtman
or  by recollections of Likhtman and others. From these papers,
and from the problems Golfand formulated for Likhtman's PhD work,
it is evident that Golfand clearly saw the contours of the theoretical
construction they were building together with Likhtman, 
posed the right questions,
and found adequate theoretical tools for their solution.
Perhaps, some weakness was on the side of phenomenology.
For instance, the issue of the parity nonconservation in the
Golfand-Likhtman model, a persistent theme  in Refs. \cite{one,three},
was never elaborated in full. This is easily explained by the
isolation in which they  were working, and  the lack of enthusiasm
 on the side of  their colleagues.
The  soil fertile to the ideas 
of Wess and Zumino, provided by CERN in 1974, was totally absent
in the case of Golfand and Likhtman.

\section*{Glimpses}

Golfand was a frequent participant of the
ITEP theory seminars.  I used to bump into him in the corridors of
ITEP regularly. At first I did not know who this
small man, with warm eyes and  a kind smile, was. So, I asked my
thesis advisor, Prof. B.L. Ioffe. Ioffe lowered his voice 
 to the level of whisper and replied that this was Golfand,
the discoverer of supersymmetry. Later, whenever he spoke of him,
Ioffe would automatically lower his voice even if we were alone 
in Ioffe's office. This would emphasize, without any words, that
Golfand was a {\em nonperson}. 

Everybody who knew Golfand remembers his smile and his eyes. 
Usually
he looked  a
little bit out of touch with reality, decoupled from the surrounding
world, with thoughts directed inside rather than
outside. 
Marinov wrote in 1994 in {\em  Proceedings of the  Israeli Physical 
Society}
that the ``Technion colleagues will remember forever 
Golfand's  smile and his quiet and sympathizing eyes.''  Elsewhere he 
elaborates\footnote{An excerpt from the interview with N. Portnova, 
1995,
unpublished.}: ``It was extremely interesting to socialize with Golfand.
He was sparing with words, he listened more than he talked; his eyes,
that were always alive and radiated warm energy, participated in the
conversation."
 
Here
is how Lars Brink describes Golfand:
``When  Mike [Green] and I gave talks at
Lebedev in 1984 he had managed to sneak in, and I met him in the 
shadow there.
The last time I saw him was at the Sakharov meeting in 1991.  He had 
emigrated
then and was back, so I talked to him several times. He represented to 
me a
character that I have only seen in Russia, the enormous warm heart, the
sadness around the eyes, somewhat subdued, a person whom you 
instantly like
and feel complete confidence in...'' Stanley Deser, who also knew Golfand 
personally, called him a man who came premature.\footnote{
I quote here from  Lars Brink and Misha Marinov not accidentally.
Lars (together with S. Deser, D. Gross, Y. Ne'eman, B. Zumino and some
other Western physicists) was absolutely instrumental in Golfand's
survival through the years of unemployment. It was their victory
when Golfand was reinstated in FIAN in 1980. 
 Misha, a recent immigrant
in Israel himself, did whatever he could to help
 Golfand to ``blend in'' into a complicated Israeli  life  during
the  most painful transition
period. Golfand's knowledge of Hebrew was rudimentary, and he would
be essentially helpless without Marinov's constant
assistance. Later Marinov took the idea of this Memorial
Volume  close to  heart; he helped 
to locate Golfand's widow, Mrs. Koretz-Golfand, in Israel.}

After I drafted this introduction,
I sent it to  a few colleagues whose opinion I value, soliciting
comments. As a result, I got a letter from
Prof. B. Ioffe which, to my mind, adds important touches to 
Golfand's human and scientific portrait. I reproduce it below,
with insignificant abbreviations. 

Ioffe writes: ``I knew Golfand  from 1951. Very close friendly relations 
developed
beginning in  around 1957: we visited each other at home, which   
is very unusual for me. Once  we celebrated 
together the New Year (1958 or 1959, I do not remember exactly), 
meeting the New Year   midnight 
in a frosty and snowy forest  (this was near
Povarovo, 50 km from Moscow) by the fire we made. We were on skis, 
Golfand liked skiing as much  as I did.

In the 1950's and 60's I often discussed physics
with him;  such discussions were very fruitful for me. 
Almost nobody knows now that Golfand  invented the path integral 
formulation in  field theory
independently from
Feynman. It was in the early 1950's (probably, in 1952). 
He represented a  field theory (he considered the scalar field theory)
as an  integral over the  mesonic fields. Golfand did not 
 follow Feynman's route who started  from path integrals in
quantum mechanics. In fact, he did not know
of Feynman's work at that time. When later he gave a talk at 
the Tamm seminar,  people were very
skeptical, because nobody understood the subject -- the presentation 
was different  from Feynman's,
 and only very few in the audience knew about 
Feynman's paper, and nobody understood it either. 
After some time it became clear, that it was just the functional integral.

When Golfand was unemployed, there was a serious problem with 
getting  permission
for Golfand's participation in ITEP seminars. As you remember, 
all ``outside''  participants were to be included in a ``list'',
which had to be cleared through the ITEP Regime Department. 
The  permission  would be  granted only
to people who had positions  in physics institutes;
the name of the institute had to be explicitly indicated
in the list we submitted to the Regime Department
before each seminar.
 Since I was responsible for the list  and 
was signing it each time (formally till 1977 it was
Berestetsky's  signature, but in fact I did it), I was committing fraud
on a regular basis  ``putting'' Golfand to some {\em ad hoc}
 institute, just to let him in.
My whisper referred not to Golfand's name  {\em per se}, but to the fact, 
that  he was unemployed. At this difficult time, we continued seeing each 
other, often exchanged  phone calls, etc.  After his divorce,
our relations cooled off a little, though. 

When I heard that Golfand
was fired from FIAN, I expressed my
dissatisfaction to a few FIAN people.  Each time the response  was:
it was  not me who  did it.
As I remember, Golfand was unemployed not all the time
from 1973 to '80; sometimes he had a part-time teaching
job at a technical college. After he was taken back to FIAN,
Golfand continued to attend the ITEP seminars
on a regular basis, but he refused to participate in the 
FIAN seminars.  That  was the demonstration of how strongly he was 
offended.''

\section*{About  This Volume}

When the idea of this  Memorial
Volume came to my mind, I wrote a letter
to my colleagues, fellow theoretical physicists.
I sent it to about two dozen active members
of the HEP community, those who determine the trends of the
modern high-energy physics, and to several
physicists from the younger generation -- some of them  I considered
to be  rising stars. The response was overwhelmingly positive.
With one or two exceptions, all agreed to participate enthusiastically,
and  I got many very valuable suggestions as to the structure
of the Volume. Its scientific part
will, hopefully, represent a full picture of the huge tree
into which supersymmetry grew today.
The book will be used in the community
in the years to come, and this is the best tribute to Golfand one can think 
of. I am sincerely grateful
to all participants of the project, to whom I would like to say thank you.
Together, we did a good job.   

Also of importance is the first part of the book,
which consists of the memoirs of Golfand's
widow, Mrs. Natasha  Koretz-Golfand, and his former student,
Dr. Evgeny Likhtman. They are both emotional and moving,
precious evidence of the past, gone forever... I am
very grateful to Mrs. Koretz-Golfand and to Dr. Likhtman
for their willingness to share with us, and with those who will 
come after us, their personal recollections. Also included
in the first part is a historical survey of the scientific ideas that
paved the way to supersymmetry, written by Prof. M.  Marinov,
and the English translation of the Golfand-Likhtman paper 
from the Tamm memorial Volume.  

\section*{Conclusions}

Golfand's career in theoretical physics spanned over 40 years.
He was the author of several dozen papers\footnote{See the list of 
Golfand's publications at the end of this Volume.} devoted to 
aspects of field theory, out of which two\footnote{In fact, one can view
these two papers as a short and long version of one and the same
paper. In essence, Golfand will be remembered as a one-work man. 
How  many  200-paper 
collections  would be gladly traded   for a single paper,  like that?} 
opened to us the
doors to the superworld, that will stay with us forever.
I will risk to say that the discovery of supersymmetry
was the single most important contribution of  Soviet 
fundamental  physics after  WW II.
I will go so far as conjecture 
that supersymmetry will play the same revolutionary role 
in  physics of the 21-st
century as  special and general relativity in 
 physics of the 20-th century.   Treatises
on the pioneers of supersymmetry -- and Yuri Golfand
definitely belongs to them --  will be written
by professional historians of science.

\vspace{0.7cm}
\noindent
Minneapolis, August 24, 1999

\end{document}